\documentclass[aps,prl,twocolumn,showpacs,preprintnumbers,amsmath,amssymb]{revtex4}   
\usepackage{graphicx}% Include figure files
\usepackage{dcolumn}% Align table columns on decimal point
\usepackage{bm}% bold math

\newcommand {\beq} {\begin{equation}}
\newcommand {\eeq} {\end{equation}}
\newcommand {\beqa}{\begin{eqnarray}}
\newcommand {\eeqa}{\end{eqnarray}}
\newcommand {\del} {\partial}
\newcommand {\tr}{{\rm tr\,}}

\newcommand {\ee}{\mbox{e}}

\begin{document}

%%%%%%%%%%%%%%%%%%%%%%%%%%%%%%%%%%%%%%%%%%%%%%%%%%%%%%%%%%%%%%%%%%%%
%  TITLE / AUTHOR                                                  %
%%%%%%%%%%%%%%%%%%%%%%%%%%%%%%%%%%%%%%%%%%%%%%%%%%%%%%%%%%%%%%%%%%%%

\title{
Schwarzschild radius from Monte Carlo calculation
%evaluation
of the Wilson loop\\
in supersymmetric matrix quantum mechanics}
 
\author{Masanori Hanada$^{1}$}
\email{masanori.hanada@weizmann.ac.il}
\author{Akitsugu Miwa$^{2,3}$}
\email{akitsugu@hep1.c.u-tokyo.ac.jp}
\author{Jun Nishimura$^{4,5}$}
\email{jnishi@post.kek.jp}
\author{Shingo Takeuchi$^{5,6}$}
\email{shingo@apctp.org}

%\address{
\affiliation{
$^{1}$Department of Particle Physics, 
Weizmann Institute of Science, 
Rehovot 76100, Israel \\
$^{2}$Institute of Physics, University of Tokyo,
Komaba, Meguro-ku, Tokyo 153-8902, Japan\\
$^{3}$Harish-Chandra Research Institute, Chhatnag Road, Jhunsi,
Allahabad 211019, India\\
$^{4}$High Energy Accelerator Research Organization (KEK), 
		Tsukuba 305-0801, Japan \\
$^{5}$Department of Particle and Nuclear Physics,
School of High Energy Accelerator Science,
Graduate University for Advanced Studies (SOKENDAI),
Tsukuba 305-0801, Japan \\
$^{6}$Asia Pacific Center for Theoretical Physics (APCTP), 
Pohang, Gyeongbuk 790-784, Korea
}

\date{November 2008; preprint: WIS/18/08-OCT-DPP, UT-Komaba/08-20, KEK-TH-1287, APCTP Pre2008-007}
%, hep-th/yymmnnn
%\today %%new

%%%%%%%%%%%%%%%%%%%%%%%%%%%%%%%%%%%%%%%%%%%%%%%%%%%%%%%%%%%%%%%%%%%%
%  ABSTRACT    	                                                   %
%%%%%%%%%%%%%%%%%%%%%%%%%%%%%%%%%%%%%%%%%%%%%%%%%%%%%%%%%%%%%%%%%%%%

\begin{abstract}
In the string/gauge duality it is important to
understand how the space-time geometry
is encoded in gauge theory observables.
%observables in gauge theory. 
We address this issue
in the case of the D0-brane system
%by considering the case
%with a stack of $N$ D0-branes 
%in type IIA superstring theory 
at finite temperature $T$.
%by explicit calculations
%in the D0-brane system at finite temperature.
Based on the duality, 
%gauge/string duality, 
the temporal Wilson loop operator $W$ in gauge theory
%encodes the information of the dual black hole geometry as
is expected to contain
the information of the Schwarzschild radius $R_{\rm Sch}$
of the dual black hole geometry as
%$\log \langle W \rangle = {R_{\rm Sch} \over 2 \pi \alpha'}{1 \over T}$.
$\log \langle W \rangle = R_{\rm Sch} / (2 \pi \alpha' T)$.
%where $R_{\rm Sch}$
% and $T_H$
% and $1/2 \pi \alpha'$ 
%is the Schwarzschild radius.
% and the Hawking temperature, respectively.
This translates to the power-law behavior
%$\log W  = 1.89 \cdot (T_H/\lambda^{1/3})^{-3/5}$,
$\log \langle W \rangle = 1.89 \cdot (T/\lambda^{1/3})^{-3/5}$,
where $\lambda$ is the 't Hooft coupling constant.
%at strong coupling, wh
We calculate the Wilson loop on the gauge theory side 
in the strongly coupled regime
%by a Monte Carlo method, which was recently used to
%extract the dual black hole thermodynamics successfully.
by performing Monte Carlo simulation of supersymmetric
matrix quantum mechanics with 16 supercharges.
%which was previously shown to reproduce the dual black hole thermodynamics.
The results 
%for the Wilson loop 
reproduce the expected power-law behavior 
%up to a subleading term, 
up to a constant shift, which
%can be identified as $\alpha '$ corrections
is explainable as $\alpha '$ corrections
on the gravity side. 
%which can be understood by considering $\alpha '$ corrections
%on the gravity side.
\end{abstract}


%11.25.Tq Gauge/string duality
%11.15.Tk Other nonperturbative techniques
%11.25.-w Theory of fundamental strings

%\pacs{11.25.-w; 11.25.Sq}
%11.25.-w Theory of fundamental strings
%11.25.Sq Nonperturbative techniques; string field theory

\maketitle 
%%%%%%%%%%%%%%%%%%%%%%%%%%%%%%%%%%%%%%%%%%%%%%%%%%%
\paragraph*{Introduction.---}%%%%%%%%%%%%%%%%%%%%%%
%%%%%%%%%%%%%%%%%%%%%%%%%%%%%%%%%%%%%%%%%%%%%%%%%%%

String/gauge duality, which originated from 
the AdS/CFT correspondence \cite{Maldacena:1997re},
% is a duality between
%superstring theory on the AdS background and large-$N$
%gauge theories
has been investigated intensively
over the past decade.
Remarkable developments that have been achieved
%so far 
include generalization to various cases, 
confirmation by explicit calculations, 
and applications to various branches of physics
such as hadron physics and condensed matter physics.
%one of the most important subjects
%in the present theoretical particle physics. 
%In the context of quantum gravity,
%Coming back to 
%In relation to 
%

From the viewpoint of 
%original motivations for 
string theory,
the duality 
%is important since it
%expected to play an important role
%imay enable us to understand 
%may 
enables us to study
%allowsin understanding 
quantum aspects of gravity
including its non-perturbative effects
from the gauge theory side, which 
%seems to be 
is more tractable. In this regard it is important 
to understand how gauge theory captures 
the information of space-time geometry \cite{emergent}.
%
%In particular, the Schwarzschild radius is one of the 
%most fundamental quantities that characterize
%the black hole geometry.
%Based on the string/gauge duality, one can show 
Based on the duality at finite temperature \cite{Witten:1998zw}, 
one can show that a temporal Wilson loop operator
% \cite{endnote}
% $W$ 
in gauge theory
% at finite temperature
%the matrix quantum mechanics 
%carries the information of the Schwarzschild radius
% $R_{\rm Sch}$ of
is related directly to
the Schwarzschild 
radius \cite{PML},
%{Rey:1998bq,Kruczenski:2005pj,Headrick:2007ca},
which is a fundamental quantity that characterizes
the dual black hole geometry. 
(See also Refs.\ \cite{Brandhuber:1998bs} for related works.)
% in a rather direct way.
In this Letter 
we confirm this prediction
by first-principle calculations
%explicit calculation 
on the gauge theory side.
%of the temporal Wilson loop in gauge theory. 
%
%we perform explicit calculations 
%of the temporal Wilson loop
%on the gauge theory side
%
% in the strongly coupled gauge theory,
%and confirm this prediction.
%Our result provides a rare opportunity to
%confirm 
%
%Note that 
%this is the first confirmation of 
%This is the first time that 
%Our result is important also in that 
%Note also that 
%it confirms
Note that this is the first confirmation of
the prescription \cite{Rey:1998ik} for
calculating the Wilson loop 
based on the string/gauge duality 
in a non-conformal theory
without protection by supersymmetry.
%for the first time 
%has been confirmed 
%by first-principle calculations
%at the 
%from first principles.
%which has been done so far only in the case 
%of half BPS Wilson loops in the superconformal case 
%with at zero temperature.
As such, we consider our results to have impact also 
in applications of the duality to 
%studies of
realistic gauge theories.
% such as QCD.

The string/gauge duality we study 
is the one \cite{Itzhaki:1998dd} associated
%corresponds to the case
%We study the case
with a stack of $N$ D0-branes 
in type IIA superstring theory
at finite temperature $T$.
The worldvolume theory of the D0-branes
is given by 1d U($N$) gauge theory or matrix quantum mechanics (MQM)
with 16 supercharges, and
% at finite temperature, in particular, 
the dual geometry
is given by the near-extremal black 
0-brane solution in type IIA supergravity.
The supersymmetric MQM
%matrix quantum mechanics
has been studied
% at various 't Hooft coupling constant
by Monte Carlo simulation \cite{Anagnostopoulos:2007fw}
in the Fourier space \cite{Hanada-Nishimura-Takeuchi}.
%for the first time by
%using the Fourier-mode simulation method 
In particular, the results for the internal energy
at various (effective) 't Hooft coupling constant
interpolated nicely the weak coupling behavior 
obtained by the high temperature expansion \cite{HTE}, 
and the strong coupling behavior predicted by
the black hole thermodynamics of the
dual geometry. (Consistent results were obtained 
also by using a
lattice approach \cite{Catterall:2008yz}.)

Here we apply this method to the calculation 
of the temporal Wilson loop operator, and demonstrate
that one can 
%its connection to 
extract the Schwarzschild radius of the dual black hole 
geometry from it.
See Ref.\ \cite{KLL}
for earlier discussions on
a similar issue
in the same model
using other observables
and other calculation techniques.
%based on different calculation techniques.
%attempts to resolve the geometry
%in the supersymmetric MQM using other observables
%and other calculation techniques.

% as
%$\log  W  =
%{R_{\rm Sch} \over 2 \pi \alpha' }{1 \over T_H}$\,
%in the strong coupling limit.
%Here $T_H$ and $1 \over 2 \pi \alpha'$ are the Hawking
%temperature and the string tension, respectively.
%% and $C$ is a numerical constant.

%%%%%%%%%%%%%%%%%%%%%%%%%%%%%%%%%%%%%%%%%%%%%%%%%%%
\paragraph*{Wilson loop in the dual string theory.---} %%%%%%%%%%%%%%
%%%%%%%%%%%%%%%%%%%%%%%%%%%%%%%%%%%%%%%%%%%%%%%%%%%

Let us review the calculation of the Wilson loop 
based on the string/gauge duality \cite{Rey:1998ik}
for general D-branes.
In addition to a stack of $N$ D-branes, which are placed
on top of each other creating a curved background geometry,
we consider a single probe D-brane, which is placed far away from them
in parallel.
%separated from them.
Since D-branes are objects which a fundamental string can 
end on, we may consider such a string stretched between 
the probe D-brane and one of the $N$ D-branes.
The amplitude for the string propagating along 
a certain loop $\mathcal{C}$\, on the D-brane
%$\Gamma$\,.
can be calculated in two different ways.
%The conjecture concerning Wilson loop is based on the
%comparison of two different 
%computations of the amplitude for a string stretching between the 
%$N$ D-branes and the probe D-brane 
%and propagating along a certain loop $\Gamma$\,.

First in the worldvolume theory of the $N$ D-branes,
the process is viewed as a heavy test particle
in the fundamental representation 
of the U($N$) group propagating along the loop.
% with mass $M$ 
The amplitude is therefore given by 
\beq
\mathcal{A} = 
\langle W(\mathcal{C}) \rangle \, 
\ee^{-M \ell}  \ ,
\label{amp_gauge}
\eeq
where $W(\mathcal{C})$ represents 
the Wilson loop associated with 
the loop $\mathcal{C}$, whose perimeter
has the length $\ell$.
%and $\ell$ represents the length of the
%its perimeter.
%$W(\Gamma)$ defined on the loop $\Gamma$\,.
The mass $M$ of the test particle, which appears in 
(\ref{amp_gauge}),
is given by the distance of the $N$ D-branes 
and the probe D-brane.

Next we view the same process on the gravity side
as a string propagating
in the curved space-time background,
which is created by the $N$ D-branes.
%the stack of $N$ D-branes.
The amplitude is calculated by the path-integral
over the worldsheet 
%configurations 
attached to the
loop $\mathcal{C}$ on the probe D-brane as
\beq
\mathcal{A}
 = \frac{1}{N}
\int_{\mathcal{C}}
%\Gamma 
{\rm e}^{-S_{\rm string}} \ ,
\label{amp_grav}
\eeq
%of the fundamental string 
%and evaluate the amplitude of a probe string attached 
%to the loop $\Gamma$ on the space-time boundary.
where $S_{\rm string}$ represents the worldsheet action,
whose bosonic part
% of $S_{\rm string}$ 
is given by the Polyakov action
\begin{equation}
S_{\rm P}
= {1 \over 4 \pi \alpha'}\!\!
\int\!\! d^2\sigma
\sqrt{h}
\Big(h^{ab} g_{MN} \partial_a x^M \partial_b x^N
+ \alpha' \phi R_{(2)} \Big) \ .
\label{Polyakov}
\end{equation}
Here $x^M$ represents the embedding of the worldsheet
into the target space with the metric
$g_{MN}$ $(M,N=1,\ldots 10)$, while
$R_{(2)}$ represents 
the two-dimensional scalar curvature
defined for the worldsheet metric $h_{ab}$ $(a,b=1,2)$\,.
The effective string coupling is given as
${\rm e}^\phi$ in terms of the dilaton field
$\phi$.  
We have omitted 
a term in (\ref{Polyakov})
depending on the NS-NS $B$-field
since the background we are considering
%that appears in our set up
does not have non-zero $B$ field.

In what follows, we will be mostly interested in 
the parameter region,
in which the string coupling is so small
that we only have to consider
% can restrict ourselves to
the disk amplitude in (\ref{amp_grav}).
We also restrict ourselves to
small $\alpha'$, which corresponds to 
a large string tension,
so that the path integral (\ref{amp_grav})
is dominated by the saddle-point configuration.
In this parameter region, one can
use the classical solution to the supergravity
as the background.
%esemi-classical analysis with $\alpha'$ as a 
%perturbation parameter.
%On the gauge theory side, 
%these restrictions correspond
%it corresponds to 
%Correspondingly on the gauge theory side, 
According to the dictionary of the string/gauge
duality, the parameter region corresponds to
%
%one has to take
taking 
the planar large-$N$ limit
with large 't Hooft coupling constant on the gauge theory side.
%On the gauge theory side, this corresponds to 
%taking the planar large-$N$ limit
%with large 't Hooft coupling.

Equating (\ref{amp_gauge}) and 
(\ref{amp_grav}), 
we obtain a formula
which relates
%enables the computation of
the Wilson loop in the strongly coupled gauge theory
to the string amplitude on the classical background geometry.
%from the gravity side.
%Then we have the following conjecture 
%\cite{Rey:1998ik}:
%\beq
%\big\langle W(\Gamma) \big\rangle = \int_\Gamma 
%{\rm e}^{-S_{\rm string}}\,.
%\label{RY-M}
%\eeq
%
The explicit check of this formula 
has been discussed
only in highly symmetric cases.
In the 4d ${\cal N}=4$ super Yang-Mills theory (SYM), 
in particular,
%in the planar large-$N$ limit
%with large 't Hooft coupling,
it is argued that the gauge theory computation 
of a circular Wilson loop, which is half BPS, 
reduces to a matrix integration with a Gaussian 
weight \cite{Erickson:2000af,Drukker:2000rr,Pestun:2007rz}.
%matrix model 
%\cite{Erickson:2000af,Drukker:2000rr,Pestun:2007rz}.
The obtained result
% for the Wilson loop 
indeed
%ing expectation value of the Wilson loop
agrees with the prediction from the gravity side.
%In the planar limit, t
The agreement can be understood also from 
the scale invariance of the worldsheet theory
near the D3-branes \cite{Kawai:2007ek}.
%in the large 't Hooft coupling.
%We are going to make an explicit check
%in more generic and less symmetric cases
%including the case at finite temperature.

%even in the large 't Hooft coupling limit
%since the gauge theory computation becomes complicated.

%%%%%%%%%%%%%%%%%%%%%%%%%%%%%%%%%%%%%%%%%%%%%%%%%%%
\paragraph*{The D0-brane case.---} %%%%%%%%%%%%%%
%%%%%%%%%%%%%%%%%%%%%%%%%%%%%%%%%%%%%%%%%%%%%%%%%%%

From now on, let us restrict ourselves to the D0-brane case.
The gauge theory side is described by the supersymmetric MQM
\begin{eqnarray}
S_{\rm SQM}
&=& 
%\frac{1}{g_{\rm YM}^2}
\frac{N}{\lambda}
\int_0^{\beta}  
d t \, 
\tr 
\bigg\{ 
\frac{1}{2} (D_t X_i)^2 - 
\frac{1}{4} [X_i , X_j]^2   
\nonumber \\
&~& 
%\left.
+ \frac{1}{2} \psi_\alpha D_t \psi_\alpha
- \frac{1}{2} \psi_\alpha (\gamma_i)_{\alpha\beta} 
 [X_i , \psi_\beta ]
\bigg\} \ , 
\label{MQM action}
\end{eqnarray}
where $D_t  = \del_t
%\frac{\partial}{\partial t} 
  - i \, [A(t), \ \cdot \ ]$ represents the covariant derivative
with the gauge field $A(t)$ being an $N\times N$ Hermitian matrix.
The model can be viewed as 
a 1d U($N$) gauge theory with adjoint matters
$X_i(t)$  $(i=1,\cdots,9)$ and $\psi_\alpha(t)$ $(\alpha=1,\cdots , 16)$,
which are bosonic and fermionic matrices, respectively.
The extent of the Euclidean time direction $\beta$ 
corresponds to the inverse temperature $\beta\equiv T^{-1}$,  
and the fermions $\psi_\alpha$ obey anti-periodic boundary conditions.
%The planar limit corresponds to taking the $N\rightarrow \infty$ 
%limit with fixed 't Hooft coupling constant $\lambda$.

We consider
%Since the gauge theory lives in 1d, let us consider
the loop $\mathcal{C}$ to be winding once 
around the temporal direction.
The corresponding Wilson loop operator 
in the gauge theory
%,which is refered to as the Polyakov-Maldacena loop
is given by \cite{Rey:1998ik}
\beq
W = {1 \over N} \, {\rm tr\,P}\!\exp
%\bigg(
\int  _0 ^{\beta} d t \Big( i A (t) + n_i X_i(t) \Big)
%\bigg) 
\ ,
\label{WM_loop}
\eeq
%where the path $\Gamma$ is taken to be the temporal circle. 
where $\vec{n}$ is a 
%9-dimensional 
unit vector
in ${\rm R}^9$
specifying the direction in which the probe D0-brane is
separated.
Note that the adjoint scalar appears in (\ref{WM_loop})
%on the exponent
unlike the definition of the Polyakov line
since the end of the string
% on the $N$ D-branes
is coupled not only to the gauge field
but also to the adjoint scalar.
% in a way depending on $\vec{n}$.
The overall factor $1/N$ is introduced to make 
the quantity finite in the planar large-$N$ limit.
%This forces us to use 
We have used the same normalization
on the right hand side of (\ref{amp_grav}).
%in order to be able to equate (\ref{amp_gauge}) and 
%(\ref{amp_grav}).

The gravity dual of the supersymmetric 
MQM
%matrix quantum mechanics
is given by the near-horizon geometry of 
the (Euclidean) near-extremal black 0-brane solution
in type IIA supergravity.
%Let us first consider the $\lambda \rightarrow \infty$ limit on the 
%gauge theory side, which corresponds to omitting
%all the $\alpha '$ corrections on the gravity side.
%Then 
%the metric of the dual geometry
In particular, the metric is given by \cite{Itzhaki:1998dd} 
\begin{equation}
{}\!\!\! {ds^2 \over \alpha'} 
\!=\!
{U^{7/2} f(U) \over  \sqrt{d_0 \lambda}}dt^2 
\!\!+\!
{ \sqrt{d_0 \lambda} \over U^{7/2} f(U)} dU^2 
\!\!+\! 
{ \sqrt{d_0 \lambda} \over U^{3/2}} d \Omega_8^2 \ ,
\label{metric1}
% \\
%{\rm e}^{\phi} & = (2 \pi)^2 g_{\rm YM}^2
%\bigg(
%{g_{\rm YM}^2 d_0 N \over U^7}
%\bigg)^{3/4}\,.\label{phi1}
\end{equation}
where $f(U) =  1  - U_0^7/U^7 $
and $d_0 \equiv 2^7 \pi^{9/2} \Gamma(7/2)$.
% is a numerical factor and 
The Schwarzschild radius and 
the inverse Hawking temperature are given by
%The inverse Hawking temperature can be obtained as
\beq
R_{\rm Sch} = \alpha' U_0 \, , \quad\quad
%T = \frac{7}{4} \pi  \frac{U_0^{5/2}}{\sqrt{d_0 \lambda}}
\beta = \frac{4}{7} \pi  \sqrt{d_0 \lambda} U_0^{-5/2} \ .
\label{betaU0}
\eeq
%by requiring that conical singularity should not appear
%at $U=U_0$ when one compactifies the Euclidean time direction $t$
%to $\beta$.
%
%compactifying the Euclidean time direction $t$
%imposing the absence of a conical singularity at $U=U_0$\,,
%
%In order to avoid a conical singularity at $U=U_0$\,,
%the Euclidean time direction $t$ has to be compactified with 
%the period 
%\beta = 4 \pi  \sqrt{d_0 \lambda}/7 U_0^{5/2}$\,.

Let us evaluate the string disk-amplitude (\ref{amp_grav})
in the background geometry (\ref{metric1}).
In the $\alpha' \rightarrow 0$ limit,
the second term in \eqref{Polyakov} 
can be omitted, and one can replace the string action by
the Nambu-Goto action $S_{\rm NG}$,
which is nothing but 
the area of the string worldsheet times the string tension.
%by solving the equation of motion
%for the worldsheet metric $h_{\alpha\beta}$
%and by substituting it in back to \eqref{Polyakov} 
%by its classical solution
%one obtains 
%one can replace the Polyakov action
%by the Nambu-Goto action $S_{\rm NG}$.
%or low temperature limit is given by 
%the classical value of \eqref{Polyakov} 
%with neglecting the second term; 
%This value is equivalent to the classical value of 
Following the proposal \cite{Rey:1998ik}, we consider a string 
worldsheet localized in the S$^8$ direction.
Then, due to the form of the metric (\ref{metric1}),
the Nambu-Goto action for the minimal area is given by
%Following the proposal \cite{Rey:1998ik}, we consider a string 
%worldsheet localized in the S$^8$ direction, 
%and assume the ansatz $t=\tau$, $U=U(\sigma)$\,
%$( \sigma_{\rm min} \leq \sigma \leq \sigma_{\rm max} )$. 
%Then the equation of motion derived from $S_{\rm NG}$
%is solved by any increasing function 
%$U(\sigma)$ satisfying $U(\sigma_{\rm min})=U_0$ and 
%$U(\sigma_{\rm max})=U_{\rm \infty}$\,.
%Here, $U_\infty$ is the IR cutoff.
%Then the classical value of the Nambu-Goto action is given by 
%% \beq
%% S_{\rm NG}= 
%% { 1 \over 2 \pi} \beta (U_\infty \! - \! U_0)
%% \ , 
%% \label{S_NG}
%% \eeq
%Here $g$ is the determinant of the induced metric.
$S_{\rm NG}= 
{ 1 \over 2 \pi} \beta (U_\infty \! - \! U_0)$,
where $U_\infty$ represents the position of the probe D0-brane.
%which plays the role of the IR cutoff.

Since the perimeter $\ell$ of the Wilson loop
in (\ref{amp_gauge}) is given by $\ell = \beta$ in 
the present set-up, we obtain the identity
\beq
\log \langle W(\mathcal{C}) \rangle 
- \beta M = 
\frac{\beta  U_0} {2 \pi} 
- 
\frac{\beta  U_\infty} {2 \pi}  \ .
\label{identityW}
\eeq
Considering that 
the mass of the test particle on the gauge theory
side is given by the position of the probe D0-brane,
it is natural to identify the second terms on both sides 
of (\ref{identityW}).
This follows also from the prescription 
proposed in Ref.\ \cite{Drukker:1999zq} based on T-duality.
Thus we obtain
\beq 
\log \langle W(\mathcal{C}) \rangle 
=
\frac{\beta  U_0} {2 \pi} 
=
{ \beta R_{\rm Sch} \over 2 \pi \alpha' }
%{1 \over {\cal T}_H}
=
%\sim
1.89 \, \left(\frac{T}{\lambda^{1/3}} \right) ^{-3/5} \ , 
\label{S_total}
\eeq
where we have used (\ref{betaU0}).

%%%%%%%%%%%%%%%%%%%%%%%%%%%%%%%%%%%%%%%%%%%%%%%%%%%
%\paragraph*{$\alpha'$ corrections.---} %%%%%%%%%%%%
\paragraph*{The range of validity.---} %%%%%%%%%%%%
%%%%%%%%%%%%%%%%%%%%%%%%%%%%%%%%%%%%%%%%%%%%%%%%%%%

%% The range of validity for 
%% the dual gravity description is known to be 
%% \cite{Itzhaki:1998dd}
%% \beq
%% N^{-10/21} \ll \frac{T}{\lambda^{1/3}} \ll 1 \ .
%% \eeq
%% We can show that the derivation of (\ref{S_total})
%% given above can be justified indeed in this parameter region.

%% The first inequality comes from the requirement that
%% the effective string coupling constant $e^{\phi}$ be
%% small, and it can be satisfied by taking the 
%% large-$N$ limit with fixed $T$ and $\lambda$.
%% The second inequality comes from the requirement that
%% the $\alpha '$ corrections be small.
%% In Monte Carlo simulations it is not easy to
%% go to very low $T$. Therefore we discuss what kind of 
%% $\alpha'$ corrections we might get.

%Let us discuss the $\alpha'$ corrections.
%Before we discuss possible corrections to (\ref{S_total}),
%it is helpful to 
Let us recall the range of validity for
the supergravity description \cite{Itzhaki:1998dd}.
By changing the target-space coordinates as 
$U=U_0 u^{2/5}$ and 
$t = \frac{2}{5}  \sqrt{d_0 \lambda} U_0^{-5/2} \tau$,
the metric \eqref{metric1} and the 
effective string coupling ${\rm e}^\phi$
% \cite{Itzhaki:1998dd}
become
%% \begin{align}
%% {ds^2 \over \alpha'} 
%% &=
%% \bigg({7 (d_0 \lambda)^{1/3}\over 4\pi uT } \bigg)^{{}\!3/5} 
%% \bigg[
%% {4 \over 25}
%% \bigg( \tilde f(u)  d \tau^2 
%% +
%% { du^2 \over \tilde f(u) }
%% \bigg)
%% +
%% d \Omega_8^2
%% \bigg]\,, \label{AdS*S}\\
%% {\rm e}^\phi &= {(2\pi)^2 \over N} 
%% \bigg(
%% {7 \lambda^{1/3} \over 4 \pi d_0^{1/7} u T }
%% \bigg)^{21/10}
%% %\times(uT)^{- 21 / 10}\,, 
%% \label{phi}
%% \end{align}
\begin{align}
{ds^2 \over \alpha'} 
&=
( d_0^{1/3} \mathcal{K})^{3/5}
%\bigg({7 (d_0 \lambda)^{1/3}\over 4\pi uT } \bigg)^{{}\!3/5} 
\bigg[
{4 \over 25}
\bigg( \tilde f(u)  d \tau^2 
+
{ du^2 \over \tilde f(u) }
\bigg)
+
d \Omega_8^2
\bigg]\,, \label{AdS*S}\\
{\rm e}^\phi &= {(2\pi)^2 \over N} 
( d_0^{-1/7} \mathcal{K})^{21/10} \ , \quad \quad
\mathcal{K} = {7 \lambda^{1/3} \over 4 \pi u T} \ ,
%\bigg(
%{7 \lambda^{1/3} \over 4 \pi d_0^{1/7} u T }
%\bigg)^{21/10}
\label{phi}
\end{align}
where $\tilde f(u) \equiv u^2(1 -u^{-14/5})$\,.
From \eqref{AdS*S}, one finds that
the geometry asymptotes at large $u$ 
to a geometry which is conformally 
equivalent to AdS$_2\times$S$^8$ \cite{Jevicki:1998yr},
and that the typical length scale of the geometry 
is given by 
$\rho \equiv (uT/\lambda^{1/3})^{-3/10}\alpha'^{1/2}$.
This scale should be much larger
than the string length $\alpha'^{1/2}$
for the $\alpha '$ corrections to the 
supergravity action to be negligible.
%description to be valid,
Hence, $uT / \lambda^{1/3} \ll 1$. 
%% If this scale is much larger than
%% the string length $\alpha'^{1/2}$\,,
%% the quantum fluctuations of the string is suppressed.
In this case,
%To see this more explicitly, we note that the first term
the first term in \eqref{Polyakov},
which is proportional to $\rho^2$,
%$(u T/\lambda^{1/3})^{-3/5} \alpha'$,
becomes large, and 
the semi-classical treatment for the string amplitude
(\ref{amp_grav}) is also justified.
%the semi-classical treatment
%for the string worldsheet is also justified.
% at $uT / \lambda^{1/3} \ll 1$.
Note, however, that 
we have introduced $U_\infty$.
%, which should be large.
Assuming that 
we only need to require
$U_\infty / U_0 $ to be large (but finite),
we may assume $u$ to be finite as well.
Then we obtain the condition $T / \lambda^{1/3} \ll 1$.

We also need to require
the effective string coupling ${\rm e}^{\phi}$ 
to be small.
From \eqref{phi}, we obtain
%this condition is given by 
$N^{-10/21} \ll T / \lambda^{1/3}$
noting that 
$u \geq 1$ in our finite temperature set-up.
%we obtain the condition $N^{-10/21} \ll T / \lambda^{1/3}$.
%In our finite temperature setting, 
%the coordinate $u$ has a minimal value $u \geq 1$.
%Therefore the condition can be satisfied 
%by $N^{-10/21} \ll T$.
%taking a sufficiently large $N$ with a fixed $T$.
%
%If this condition is not met, 
%we need to consider the eleven-dimensional 
%supergravity.

%%%%%%%%%%%%%%%%%%%%%%%%%%%%%%%%%%%%%%%%%%%%%%%%%%%
\paragraph*{$\alpha'$ corrections.---} %%%%%%%%%%%%
%%%%%%%%%%%%%%%%%%%%%%%%%%%%%%%%%%%%%%%%%%%%%%%%%%%

%In actual Monte Carlo simulation, 
%it is not easy to lower the temperature.
%Therefore, 

Let us discuss possible subleading terms in (\ref{S_total}) 
%at higher orders in $T/ \lambda^{1/3}$.
%This corresponds to considering 
due to $\alpha'$ corrections on the gravity side.
There are three effects one should consider:
(I) the coupling with the background $\phi$ field
represented by the second term in \eqref{Polyakov},
(II) $\alpha'$-corrections to the background fields
that appear in
%which gives rise to $\alpha'$-corrections 
%for 
the action \eqref{Polyakov}, and  
(III) the quantum fluctuation of the string worldsheet
including fermionic degrees of freedom 
in evaluating (\ref{amp_grav}).
In order to discuss the next-leading order terms,
we can treat
%discuss 
each of these effects separately.
%sgincorrection of the probe string itself, 
%which should be calculated by taking account of (I) and (II).

The effect (I) yields a constant term 
and a logarithmic term with respect to $T/\lambda^{1/3}$
in (\ref{S_total}) as one can see from 
\eqref{Polyakov} and \eqref{phi}.
The constant term includes $\log N$,
but this is canceled by the prefactor $1/N$ 
in (\ref{amp_grav}) as it should.
The effect (II) can be neglected at this order
%yields a term $(T/\lambda^{1/3})^{9/5}$,
since $\alpha'$-corrections to the type 
IIA supergravity action starts only at the ${\alpha'}^3$ 
order \cite{GW}. 
The effect (III) yields a constant term
%${\alpha'}^0$ term, i.e., $T^0$ term 
to (\ref{S_total}).
%Although it is possible that 
%the correction (III) vanishes because of symmetry,
%we assume the largest possible correction.
This effect is discussed also in the case of 
D3-branes \cite{Greensite:1999jw}. 
%, and it is shown that the correction starts generically
%at the ${\alpha'}^0$ order \cite{Greensite:1999jw}. 
In fact 
%As is discussed in that case 
%one may also get
%In fact the effect (III) may also yield
%give rise to
a logarithmic term can appear from it as well
%with respect to $T/\lambda^{1/3}$ which originates 
due to the 
%normalization of 
insertion of the ghost zero mode \cite{Drukker:2000rr}.
% \cite{Drukker:2000rr}.

%% In summary the gravity analysis
%% predicts the following behavior 
%% of the expectation value of the Wilson loop:
%% \beq
%% \log \big \langle W\big \rangle
%% = 1.89 T^{-3/5}- C
%% + {\cal O} \big(T^{3/5}\big)\,,
%% \label{Wilson_gravity side}
%% \eeq
%% where $C$ is a constant.
%% We have neglected possible logarithmic terms explained above,
%% since it is difficult to distinguish a logarithmic term from a constant 
%% by using the numerical data.
%% To evaluate the explicit value $C$
%% from the gravity side is beyond the scope of the 
%% present letter.
%% In next section, we check the behavior 
%% by using $C$ as a free parameter.

%%%%%%%%%%%%%%%%%%%%%%%%%%%%%%%%%%%%%%%%%%%%%%%%%%%
\paragraph*{Monte Carlo simulation.---} %%%%%%%%%%%%%%%%
%%%%%%%%%%%%%%%%%%%%%%%%%%%%%%%%%%%%%%%%%%%%%%%%%%%

We perform Monte Carlo simulation of the model
(\ref{MQM action}) and calculate the temporal Wilson loop
(\ref{WM_loop}) to check the prediction (\ref{S_total}).
We use the Fourier-mode simulation method 
\cite{Hanada-Nishimura-Takeuchi},
in which we take the static diagonal gauge 
$A(t) = \frac{1}{\beta} {\rm diag}
(\alpha_1 , \cdots , \alpha_N)$
with $- \pi < \alpha_a \le \pi $,
and introduce a cutoff $\Lambda$ on the Fourier modes
as $X_i (t) = \sum_{n=-\Lambda}^{\Lambda} 
\tilde{X}_{i n} \ee^{i \omega n t}$, where
$\omega = \frac{2 \pi }{\beta}$.
%Unlike the lattice approach \cite{Catterall:2008yz,Catterall:2007fp},
Supersymmetry at $T=0$, which is broken only due to 
finite $\Lambda$, is shown to recover rapidly 
as $\Lambda\to\infty$
in a simpler model \cite{Hanada-Nishimura-Takeuchi}.
The effective 't Hooft coupling constant is given 
by $\lambda_{\rm eff} = \lambda/T^3$.
In actual simulation we set
% the 't Hooft coupling constant to unity 
$\lambda=1$ without loss of generality,
so that high/low $T$ corresponds to weak/strong coupling,
respectively.

Integration over the fermionic matrices
yields a Pfaffian ${\rm Pf}{\cal M}$,
which is complex in general.
%one lowers the temperature.
According to the standard reweighting method,
one uses $|{\rm Pf}{\cal M}|$ to generate
configurations, and includes
the effect of the phase when one calculates 
the expectation values.
In fact ${\rm Pf}{\cal M}$ is almost real positive 
at sufficiently high $T$, but the fluctuation of the phase 
becomes larger as $T$ decreases, which causes the
so-called sign problem.
It turned out, however,  
that the results of the reweighting method in the temperature
regime where the sign problem is not so severe
are actually in good agreement
with what we obtain by simply neglecting the phase.
We interpret this as an effect of the large-$N$ limit,
in which the fluctuations of single trace observables 
vanish. 
%We therefore present results obtained
%by simply neglecting the phase.
For the same reason, it is expected that 
$\log \langle |W| \rangle$ agrees with
$\langle \log |W| \rangle$ in the large-$N$ limit.
We therefore calculate the latter in an ensemble 
generated with $|{\rm Pf}{\cal M}|$.
%$\langle \log W \rangle$
%to obtain better statistics.
Complete justification of these simplifications is
left for future investigations.

%% We perform simulations with $|{\rm Pf}{\cal M}|$.
%% The effect of the phase can be included by the
%% so-called reweighting method when we calculate expectation
%% values of observables.
%% We have performed such analysis in cases where the
%% phase fluctuation is not large enough to wash away
%% all the signal. But we find that the result does not
%% change within errorbars 
%% from what we obtain by simply neglecting the phase.
%% We interpret this as an effect of the large-$N$ limit,
%% in which fluctuations of a single trace observables 
%% vanish. We therefore present results obtained
%% by simply neglecting the phase.
%% For the same reason, it is expected that 
%% $\log \langle |W| \rangle$ agrees with
%% $\langle \log |W| \rangle$ in the large $N$ limit.
%% We therefore calculate the latter
%% %$\langle \log W \rangle$
%% to obtain better statistics.
%% Complete justification of these simplifications is
%% left for future investigations.

We evaluate (\ref{WM_loop}) as a limit
$W=\lim_{\nu\to\infty} W_\nu$, where
\beq
W_\nu=
%\lim_{\nu \to\infty} 
{1\over N}\tr \prod_{k=0}^{\nu-1}
\left[ 1+ \frac{\beta}{\nu} 
\Bigl\{ i A+n_i X_i(t_k) \Bigr\} \right] 
\eeq
with $t_k=\frac{k}{\nu}\beta$.
The matrices $X_i(t_k)$ are obtained
as the inverse Fourier transform
of the configurations generated by our simulation.
%% can be easily obtained from
%% configurations in the Fourier space
%% generated by our simulation.
%
%by applying the 
%inverse Fourier transformation to 
%the configurations in the Fourier 
%generated by our simulation.
Using the asymptotic behavior
$W_\nu\simeq W+\frac{\rm const.}{\nu}$
at large $\nu$, we can make a reliable extrapolation
to $\nu=\infty$.
As the unit vector $\vec{n}$, we have used
the ones
%unit vectors 
in all 9 directions with plus or
minus sign in front, and averaged over them to
increase statistics.

\begin{figure}[htb]
\begin{center}
%\rotatebox{-90}{
\includegraphics[height=6cm]{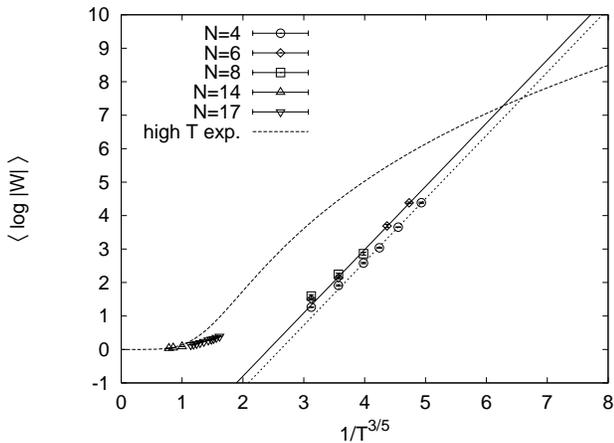}
%}
%\includegraphics[height=8cm]{maldLoopForPaper_tentative.eps}
\end{center}
\caption{
The plot of
%expectation value 
$\langle\log|W|\rangle$ for $\lambda=1$
%is plotted 
against $T^{-3/5}$. 
The cutoff $\Lambda$ is chosen as follows: 
$\Lambda=12$ for $N=4$; $\Lambda=0.6/T$ for $N=6,8$; 
$\Lambda=4$ for $N=14$; $\Lambda=6$ for $N=17$.
The dashed line represents the results
of the high-temperature expansion
up to the next-leading order
with extrapolations to
$N=\infty$, which are obtained by applying the method 
in Ref.\  \cite{HTE}.
The solid line and the dotted line represent fits
for $N=6$ and $N=4$ respectively,
to straight lines with the slope 1.89 
predicted from the gravity side at the leading order.
}
\label{fig:logW}
\end{figure}

In Fig.\ \ref{fig:logW} we plot
$\langle\log|W|\rangle$ against $T^{-3/5}$.
As $T$ decreases (to the right on the horizontal axis),
the data show a clear linear growth with a slope
consistent with the value $1.89$ predicted
in (\ref{S_total}).
In fact we can fit our data to
$\langle\log|W|\rangle = 1.89 T^{-3/5} - C$,
where $C=4.95$ for $N=4$ and $C=4.58$ for $N=6$.
%It is interesting that the power and the coefficient
%agrees with the prediction (\ref{S_total}) even at $N=4$,
%while the constant $C$ has certain $N$ dependence.
The data points for $N=8$ are very close to those for $N=6$.
%% Note, however, that the data points for $N=8$ are 
%% very close to those for $N=6$, from which we suspect that
%% $C=4.58$ is already close to the large-$N$ limit.
Note that the constant term and the logarithmic term
predicted from the gravity side are difficult to distinguish
numerically.
We therefore consider that the value of $C$ extracted above
actually represents the sum of the two terms
at the temperature regime investigated.

%%%%%%%%%%%%%%%%%%%%%%%%%%%%%%%%%%%%%%%%%%%%%%%%%%%
\paragraph*{Summary.---} %%%%%%%%%%%%%%%%%%%%%%%%%%
%%%%%%%%%%%%%%%%%%%%%%%%%%%%%%%%%%%%%%%%%%%%%%%%%%%

We have presented the first Monte Carlo
calculations of the Wilson loop
in a supersymmetric gauge theory 
at strong coupling.
Up to subleading terms anticipated from the analysis
on the gravity side,
our results are in precise agreement 
with the prediction
% (\ref{S_total}) 
from the dual supergravity. 
This is a new and highly nontrivial evidence
for the string/gauge duality. 
%Let us also recall that the derivation
%of (\ref{S_total}) involved a subtlety in
%has been derived 
%by using the standard treatment \cite{Drukker:1999zq} 
%for the IR cutoff.
%Our results suggest that this prescription is indeed correct.
%
%% %type IIA supergravity. 
%% (\ref{WM_loop}) in maximally supersymmetric 
%% matrix quantum mechanics at finite temperature. 
%% %In the gravity side we assumed the prescription 
%% %of \cite{Drukker:1999zq} as the regularization sheme.
%% We evaluated the expectation value by  
%% the Monte Carlo simulation and found beautiful agreement 
%% with the prediction (\ref{Wilson_gravity side}) from the dual 
%% type IIA supergravity. 
It would be nice to obtain the subleading terms
explicitly from the gravity side,
which
% and compare them with our data. That will provide
will provide
a nontrivial check of the duality 
including $\alpha '$ corrections.
It is also interesting to extend this work to
$\mathcal{N}=4$ SYM
%super Yang-Mills theory 
on ${\rm R}\times {\rm S}^3$,
which is possible by using the equivalence \cite{Ishii:2008ib}
in the planar limit
between the SYM
%this theory 
and a mass-deformed MQM
% with mass deformation preserving supersymmetry 
around a multi-fuzzy-sphere background.
The equivalence is confirmed by explicit calculations
at weak coupling \cite{SYMcheck}.

The fact that we were able to see
the Schwarzschild radius of the dual black hole geometry
by simulating large-$N$ matrices gives us
strong support and a firm ground
for using matrix model simulations
to study quantum gravity \cite{Monte}. 
Note that the gauge theory description is valid also
at small $\lambda$ and small $N$,
where the dual supergravity description is no longer valid.
Of particular interest is to
%go beyond the planar limit, and 
study the parameter region corresponding to M-theory.

%% Note that the gauge theory description is valid also
%% at small $\lambda$ and small $N$,
%% where the dual supergravity description is no longer valid.
%% It is most interesting to study 
%% When $N^{-10/21} \ll \frac{T}{\lambda^{1/3}} $ is not
%% satisfied, we should start to see the effects of 
%% the string coupling constant, which are described by 
%% the M-theory.
%% Further studies along this direction is very important.

%% However at further lower temperature the deviation,   
%% which is expected to describe M-theory, will emerge.  
%% Further study along this direction is very important 
%% for understanding M-theory, whose entirety is still unknown.  

%% Further study along this direction is very important 
%% for understanding M-theory, whose entirety is still unknown.  

%% In the temperature region we studied we did not find 
%% any deviation from planar limit. 
%% However at further lower temperature the deviation,   
%% which is expected to describe M-theory, will emerge.  
%% Further study along this direction is very important 
%% for understanding M-theory, whose entirety is still unknown.  

%%%%%%%%%%%%%%%%%%%%%%%%%%%%%%%%%%%%%%%%%%%%%%%%%%%%%%%%%%%%%%%%%%%%
%  ACKNOWLEDGEMENTS                                                %
%%%%%%%%%%%%%%%%%%%%%%%%%%%%%%%%%%%%%%%%%%%%%%%%%%%%%%%%%%%%%%%%%%%%

\paragraph*{Acknowledgments.---}
%We are grateful to 
The authors would like to thank
O.~Aharony, K.~N.~Anagnostopoulos, Y.~Hyakutake, 
H.~Kawai, Y.~Kazama, H.~Ooguri,
J.~Sonnenschein, A.~Tsuchiya and T.~Yoneya for discussions. 
%M.~H. is grateful to the Niels Bohr Institute,  
%the Albert Einstein Institute Potsdam, 
%University of Tokyo Hongo and RIKEN Nishina Center   
%for hospitality during his stay.
The computations were carried out on supercomputers
SR11000 at KEK
% and SX8 at RCNP)
as well as on PC clusters at KEK and Yukawa Institute. 
The work of A.\ M.\ is supported 
%in part 
by JSPS.
%% by JSPS Research Fellowships 
%% for Young Scientists.
The work of J.\ N.\ is supported 
%in part 
by Grant-in-Aid for Scientific
Research (Nos.\ 19340066 and 20540286).
%from the Ministry 
%of Education, Culture, Sports, Science and Technology.

%%%%%%%%%%%%%%%%%%%%%%%%%%%%%%%%%%%%%%%%%%%%%%%%%%%%%%%%%%%%%%%%%%%%
%  REFFERENCE                                                      %
%%%%%%%%%%%%%%%%%%%%%%%%%%%%%%%%%%%%%%%%%%%%%%%%%%%%%%%%%%%%%%%%%%%%

%\end{references}

\end{document}